\documentclass[sigconf,screen]{acmart}
\AtBeginDocument{%
  \providecommand\BibTeX{{%
    \normalfont B\kern-0.5em{\scshape i\kern-0.25em b}\kern-0.8em\TeX}}}

\newcommand{\para}[1]{\vspace{2pt}\noindent\textbf{#1.~}}
\newcommand{\ignore}[1]{}
\newcommand{\system}{\sloppy{SmartState\@}}
\newcommand{\bugname}{\sloppy{SRV\@}}

\newcommand{\eosafe}{\textsc{EOSafe}}

\definecolor{verylightgray}{rgb}{.97,.97,.97}

\usepackage{balance}
\usepackage{listings, xcolor}
\usepackage{pifont}
\usepackage{ulem}
\usepackage{array,framed}
\usepackage{setspace}
\usepackage{latexsym,fancyhdr,url}

\usepackage{enumerate}
\usepackage{xspace}
\usepackage{multirow}
\usepackage{appendix}
\usepackage{microtype}

\lstdefinelanguage{Solidity}{
  keywords=[1]{anonymous, assembly, assert, balance, break, call, callcode, case, catch, class, constant, continue, constructor, contract, debugger, default, delegatecall, delete, do, else, emit, event, experimental, export, external, false, finally, for, function, gas, if, implements, import, in, indexed, instanceof, interface, internal, is, length, library, log0, log1, log2, log3, log4, memory, modifier, new, payable, pragma, private, protected, public, pure, push, require, return, returns, revert, selfdestruct, send, solidity, storage, struct, suicide, super, switch, then, this, throw, transfer, true, try, typeof, using, value, view, while, with, addmod, ecrecover, keccak256, mulmod, ripemd160, sha256, sha3}, 
  keywordstyle=[1]\color{blue}\bfseries,
  keywords=[2]{address, bool, byte, bytes, bytes1, bytes2, bytes3, bytes4, bytes5, bytes6, bytes7, bytes8, bytes9, bytes10, bytes11, bytes12, bytes13, bytes14, bytes15, bytes16, bytes17, bytes18, bytes19, bytes20, bytes21, bytes22, bytes23, bytes24, bytes25, bytes26, bytes27, bytes28, bytes29, bytes30, bytes31, bytes32, enum, int, int8, int16, int24, int32, int40, int48, int56, int64, int72, int80, int88, int96, int104, int112, int120, int128, int136, int144, int152, int160, int168, int176, int184, int192, int200, int208, int216, int224, int232, int240, int248, int256, mapping, string, uint, uint8, uint16, uint24, uint32, uint40, uint48, uint56, uint64, uint72, uint80, uint88, uint96, uint104, uint112, uint120, uint128, uint136, uint144, uint152, uint160, uint168, uint176, uint184, uint192, uint200, uint208, uint216, uint224, uint232, uint240, uint248, uint256, var, void, ether, finney, szabo, wei, days, hours, minutes, seconds, weeks, years},  
  keywordstyle=[2]\color{teal}\bfseries,
  keywords=[3]{block, blockhash, coinbase, difficulty, gaslimit, number, timestamp, msg, data, gas, sender, sig, value, now, tx, gasprice, origin},  
  keywordstyle=[3]\color{violet}\bfseries,
  identifierstyle=\color{black},
  sensitive=false,
  comment=[l]{//},
  morecomment=[s]{/*}{*/},
  commentstyle=\color{gray}\ttfamily,
  stringstyle=\color{red}\ttfamily,
  morestring=[b]',
  morestring=[b]"
}
\setcopyright{acmlicensed}
\acmPrice{15.00}
\acmDOI{10.1145/3597926.3598111}
\acmYear{2023}
\copyrightyear{2023}
\acmSubmissionID{issta23main-p265-p}
\acmISBN{979-8-4007-0221-1/23/07}
\acmConference[ISSTA '23]{Proceedings of the 32nd ACM SIGSOFT International Symposium on Software Testing and Analysis}{July 17--21, 2023}{Seattle, WA, USA}
\acmBooktitle{Proceedings of the 32nd ACM SIGSOFT International Symposium on Software Testing and Analysis (ISSTA '23), July 17--21, 2023, Seattle, WA, USA}
\received{2023-02-16}
\received[accepted]{2023-05-03}

\begin{document}

\title{SmartState: Detecting State-Reverting Vulnerabilities in Smart Contracts via Fine-Grained State-Dependency Analysis}

\author{Zeqin Liao}
\affiliation{%
   \institution{Sun Yat-sen University}
  \city{Guangzhou}
  \country{China}
}
\email{liaozq8@mail2.sysu.edu.cn}

\author{Sicheng Hao}
\affiliation{%
  \institution{Sun Yat-sen University}
  \city{Guangzhou}
  \country{China}}
\email{haosch@mail2.sysu.edu.cn}

\author{Yuhong Nan}
\authornote{corresponding author}
\affiliation{%
  \institution{Sun Yat-sen University}
  \city{Guangzhou}
  \country{China}}
\email{nanyh@mail.sysu.edu.cn}

\author{Zibin Zheng}

\affiliation{%
  \institution{Sun Yat-sen University}
  \city{Guangzhou}
  \country{China}}
\email{zhzibin@mail.sysu.edu.cn}

\begin{abstract}

Smart contracts written in Solidity are widely used in different blockchain platforms such as Ethereum, TRON and BNB Chain. One of the unique designs in Solidity smart contracts is its state-reverting mechanism for error handling and access control. 
Unfortunately, a number of recent security incidents showed that adversaries also utilize this mechanism to manipulate critical states of smart contracts, and hence, bring security consequences such as illegal profit-gain and Deny-of-Service (DoS). In this paper, we call such vulnerabilities as the State-reverting Vulnerability (SRV). Automatically identifying SRVs poses unique challenges, as it requires an in-depth analysis and understanding of the state-dependency relations in smart contracts. 

This paper presents \system{}, a new framework for detecting state-reverting vulnerability in Solidity smart contracts via fine-grained state-dependency analysis. \system{} integrates a set of novel mechanisms to ensure its effectiveness. Particularly, \system{} extracts state dependencies from both contract bytecode and historical transactions. Both of them are critical for inferring dependencies related to SRVs. Further, \system{} models the generic patterns of SRVs (i.e., profit-gain and DoS) as SRV indicators, and hence effectively identify SRVs based on the constructed state-dependency graph.
To evaluate \system{}, we manually annotated a ground-truth dataset which contains 91 SRVs in the real world. Evaluation results showed that \system{} achieves a precision of 87.23\% and a recall of 89.13\%. 
In addition, \system{} successfully identifies 406 new SRVs from 47,351 real-world smart contracts. 11 of these SRVs are from popular smart contracts with high transaction amounts (i.e., top 2000). In total, our reported SRVs affect a total amount of digital assets worth 428,600 USD.
\end{abstract}



\maketitle

\section{Introduction}
\label{sec:intro}

Smart contracts are a specific type of program running on the blockchain system. Solidity, one of the most popular programming languages for writing smart contracts, is widely used in major blockchain platforms such as Ethereum~\cite{Ethereum}, TRON~\cite{Tron}, and BNB Chain~\cite{BNB}. Smart contracts are now supporting a wide range of Decentralized applications (DApps) in blockchain, such as decentralized finance (DeFi), decentralized gaming (GameFi), and Non-Fungible-Token (NFT). Since most blockchain systems hold a considerable value of digital assets, the security of smart contracts is of great importance to both smart contract owners (e.g., DApp developers) and users. For example, in the well-known DAO attack~\cite{luu2016making}, the attacker exploited a smart contract vulnerability (i.e., reentrancy) and caused an economic loss of 60 million USD.

\para{State-reverting and its security implications} In Solidity, state-reverting is a specific mechanism to support error handling and access control~\cite{soliditydocumentation}. More specifically, if an unsatisfied condition meets in the middle of a transaction, state-reverting allows the state variables in a smart contract to rollback to their original states.  
Unfortunately, a number of recent security incidents~\cite{GameFi, Fomo3D} showed that, smart contracts might contain vulnerable code patterns when implementing state-reverting, allowing adversaries to utilize this mechanism and manipulate certain critical states by fabricating transaction errors. Such attacks can cause severe consequences to the victim smart contracts, such as illegal profit-gain~\cite{he2021eosafe} and Deny-of-Service (DoS)~\cite{feist2019slither}. In our research, we refer to such vulnerabilities in smart contracts as the State-reverting Vulnerability (SRV).

Given the severe impact of SRVs, there has been limited research aiming at SRVs. Specifically, {\eosafe}~\cite{he2021eosafe} and WASAI~\cite{chen2022wasai} focus on detecting the rollback vulnerability caused by state-reverting through symbolic analysis and fuzzing, respectively. 
Besides, eTainter~\cite{ghaleb2022etainter} and Madmax~\cite{grech2018madmax} detect Deny-of-Service (DoS) vulnerabilities caused by state-reverting based on abnormal gas consumption. These approaches can only cover a subtype of SRV (i.e., profit-gain or DoS, see more discussion on Section~\ref{sec:ploblemstatement}). In addition, both {\eosafe}~\cite{he2021eosafe} and WASAI~\cite{chen2022wasai} are designed for smart contracts in the WASM language. Since their detection heuristics rely on language-specific features (i.e., the inline mechanism in WASM), these approaches are not applicable to Solidity smart contracts.

\para{Our work} In this paper, to fill the above research gap, we propose \system{}, a new framework for detecting state-reverting vulnerability in Solidity smart contracts. To the best of our knowledge, \system{} is the first of its kind to support SRV detection in a generic manner at the bytecode level. 

The key challenge in this research is to construct a fine-grained state-dependency graph, which is mandatory for determining which state could be affected or manipulated by an adversary from the attack surfaces (e.g., function calls from other contracts). However, state-of-the-art approaches can only infer limited state dependency information, which is not sufficient for detecting SRVs. For example, \textsc{Sailfish}~\cite{bose2022sailfish} only covers certain explicit dependencies (i.e., state Read and Write (R\&W) dependency and control flow dependency) for detecting state inconsistency bugs in smart contracts.

To effectively detect SRVs, \system{} identifies two types of state-dependency relations which are missed by prior research, namely, assertion-related state dependency (ASD) and temporal-order state dependency (TSD). Here, ASD refers to explicit dependencies related to statements such as \texttt{assert}, \texttt{revert}, and \texttt{require} for access control, while TSD refers to dependencies affected by transactions (i.e., function invocations) which is unique to smart contracts. To this end, \system{} infers such important dependencies by analyzing both the smart contract bytecode, as well as their corresponding historical transactions (see Section~\ref{sec:explicit} and~\ref{subsec:model-extraction} for more details). In this way, \system{} combines the extracted dependencies and constructs a fine-grained state-dependency graph for SRV detection. Further, \system{} models the generic patterns of SRVs (i.e., profit-gain and DoS) as SRV indicators. For example, \system{} detects whether a critical state in the SDG lacks appropriate access control checks. Therefore, it effectively identifies SRVs based on the constructed state-dependency graph.


To evaluate \system{}, we have constructed a manually-labeled dataset which consists of 91 SRVs collected from public reports and manual investigation over the top-popular smart contracts. Our evaluation results showed that \system{} achieves a precision of 87.23\% and recall of 89.13\%, indicating its high effectiveness in detecting SRVs.
In addition, by running \system{} over 2,000 popular smart contracts in the real world, our research successfully identified 11 new SRVs which have not been identified by previous research. 
The total assets affected by these SRVs reached 428,600 USD.

In summary, this paper makes the following contributions.

\begin{itemize}

\item We propose \system{}, a novel framework for detecting state-reverting vulnerabilities based on static analysis. To the best of our knowledge, \system{} is the first of its kind to detect SRVs in a generic manner.

\item We propose a set of new mechanisms (i.e., assertion-related dependency and transaction-order dependency) to construct fine-grained state dependency graph in smart contracts. 

\item We perform extensive evaluation to show the effectiveness of \system{}. In addition, by performing a large-scale study over 47,351 smart contracts in the wild, \system{} identified 406 new SRVs in the real world.

\item We release the artifact of \system{}, as well as the manual-labeled SRV dataset~\footnote{https://github.com/InPlusLab/SmartState}.

\end{itemize}

\section{Background and Motivation}
\label{sec:background}

In this section, we first lay out some basic background knowledge about smart contracts and contract states. Then, we present the problem statement and the motivation of our research. 

\subsection{Smart Contract and Contract State}
\label{sect:contractstate}

Smart contracts are a specific type of program running on the blockchain. At present, most smart contracts are written in Solidity~\cite{soliditydocumentation}. With the Solidity source code, a smart contract is then compiled into the bytecode and executed on the Virtual Machine (VM) of different blockchain platforms (e.g., Etherum~\cite{Ethereum}, TRON~\cite{Tron}, and BNB Chain~\cite{BNB}).
Similar to other program languages such as Python and Java, smart contracts consist of a set of functions and variables. Functions can be invoked through calls from smart contracts and user accounts. Such a function invocation is also known as \textit{transactions}.

Contract states are persistent data stored and accessed via global variables (i.e., state variables) in smart contracts. A transaction (function invocation) actually changes the state(s) of related smart contracts, and this process is permanently recorded on blockchain~\cite{soliditydocumentation}. 
Due to the limited storage space on the blockchain, smart contracts use state variables to store critical data, such as the owner's address, users' token balance, etc. Therefore, if an adversary could manipulate these critical states, it may bring severe negative impacts (e.g., financial losses) to the contract owner. For example, in the well-known Fomo3D attack, the attacker leverages a \textit{DoS} vulnerability to stop the purchase of other contract users (i.e., roll back the modification on the state variable that represents purchase), which caused an economic loss of 43 million USD~\cite{Fomo3D}.

\para{State-reverting mechanism} State-reverting is a unique mechanism for error handling and access control in smart contracts. If an unsatisfied condition meets in the middle of a transaction, all contract states related to this transaction can be rolled back to their previous values before the transaction. Such a state-reverting mechanism is essential for the scenario of smart contracts, as it ensures the atomicity of transactions in smart contracts. In smart contracts, state-reverting can be implemented by invoking the assertion statement (i.e., \texttt{require}, \texttt{assert}, and \texttt{revert}). For example, a smart contract with \textit{"require(tx.origin = msg.sender)"} indicates that the function can only be invoked by an external owned account (EOA) ~\cite{EOA}, if the caller of the function is not an EOA (e.g., contract account), all other state changes made in this transaction will be rolled back.



\subsection{Problem Statement}
\label{sec:ploblemstatement}

In this paper, our research focuses on vulnerabilities related to the state-reverting mechanism. We call this type of vulnerability as the State-reverting vulnerability (SRV). 

\para{State-reverting vulnerability (SRV)} Recent studies and reports~\cite{he2021eosafe} showed that the state-reverting mechanism is frequently utilized by attackers and causes severe attacks and security incidents. More specifically, in these attacks, the adversary uses the state-reverting mechanism and makes the transaction fall back, if the result of the transaction is not as expected (e.g., benefits the attacker). State-reverting attacks are more prevalent in GameFi markets, affecting game fairness and bringing heavy financial losses to contract owners~\cite{chen2022wasai} or GameFi users.

\begin{figure}[t]
\begin{lstlisting}
contract TokenGame {
  mapping(address => uint256) public SheepToken;
  mapping(address => uint256) public WolfToken;
  function MintToken(address account) public {
    uint256 seed= (random () >> 245) % 10; 
    //A random number determines gambling results 
    if ( seed != 0) { 
      SheepToken[account]++;} 
    else{
      //The state variable manipulated by attacker
      WolfToken[account]++;}   
  }    
}

contract Attacker {
  function onlyWolf(TokenGame target, tokenId) public{
    unit256 Before = WolfToken.balanceOf(address(this));
    target.MintToken(tokenId);
    unit256 After = WolfToken.balanceOf(address(this));
    // Reverting the unexpected gambling result
    require(After > Before); }
}

\end{lstlisting}

\caption{An example of state-reverting vulnerability and how attacker exploits it for profit-gain.}
\label{illustrateexample}
\end{figure}

\para{Motivating example} Figure~\ref{illustrateexample} shows an example of smart contract with state-reverting vulnerability which affects game fairness, as well as how the attacker obtains illegal profits by manipulating contract states. 
In this example, the state variables, \textit{SheepToken} and \textit{WolfToken} are two types of tokens with different values (i.e., \textit{WolfToken} is more expensive than sheep). A random number (seed) determines the gambling result, with a 90\% chance to get a \textit{SheepToken} and a 10\% chance to get the \textit{WolfToken} (line 7-11). 
Unfortunately, due to the lack of appropriate access control, an adversary can arbitrarily check the balance of these tokens before and after the gambling game (line 17-19). To maximize gain in the game, the adversary uses the require statement (line 21) to revert the whole transaction (i.e., \textit{onlyWolf}), if he is rewarded with the low-value \textit{SheepToken}. In this way, the attacker ensures he always gets the \textit{WolfToken} and brakes the game fairness with more profits.

\begin{figure}[t]
\centering
	\includegraphics[width=2.5in]{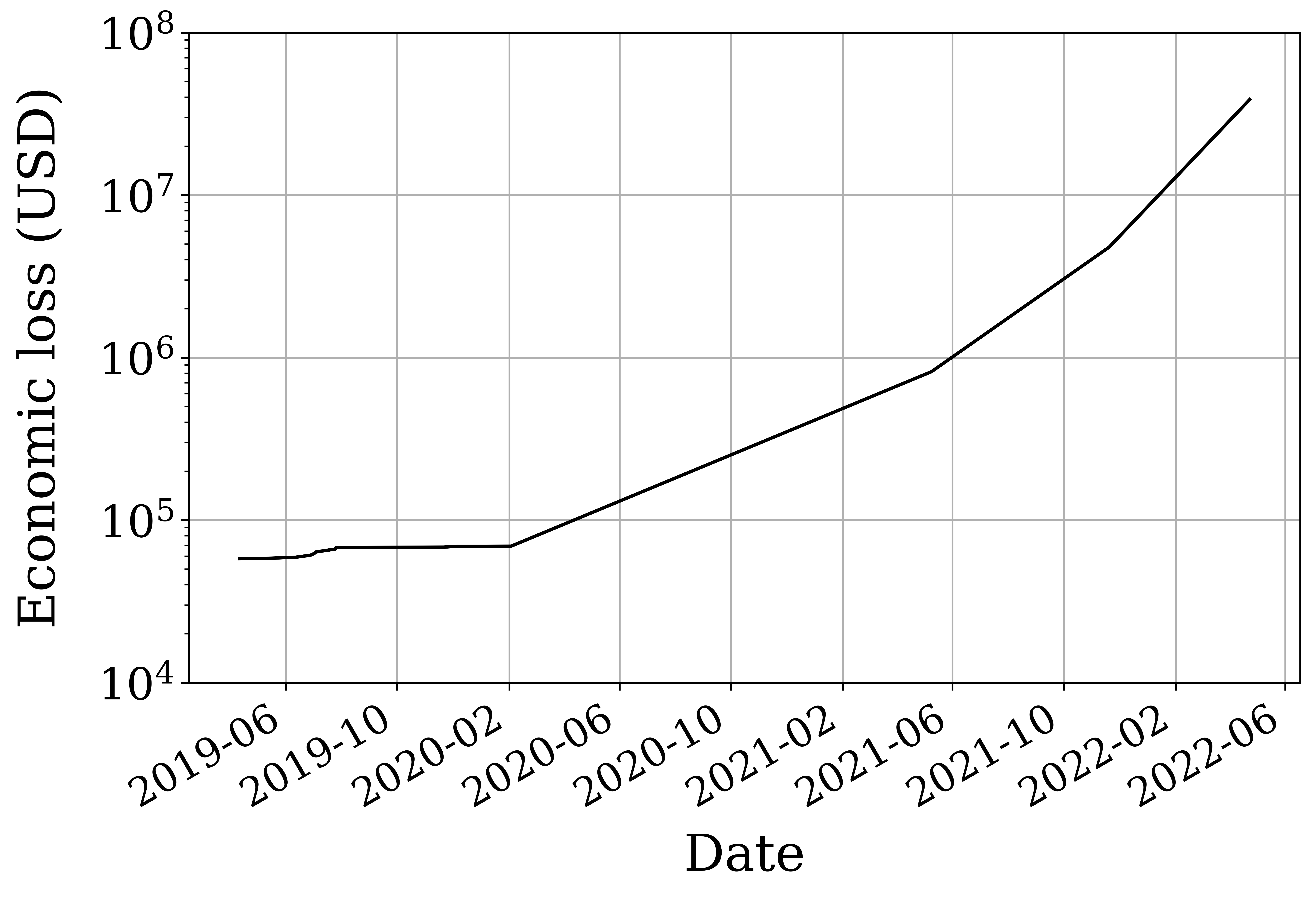}
    \caption{Summarized Economic loss caused by state-reverting vulnerability in recent three years. The sources of the collected incidents are summarized in \cite{attackscollection}.} 
    \label{economic}
    
\end{figure}

Figure~\ref{economic} summarizes the accumulated financial losses caused by state-reverting attacks in recent three years. The sources of such statistics and corresponding incidents are listed in ~\cite{attackscollection}. As can be seen, the economic loss caused by SRVs is increasing rapidly and has reached 38.69 million USD as of April 2022.

\para{Prior research and their limitations} Despite reported attack incidents caused by SRVs~\cite{Fomo3D}, there have been limited prior works on identifying SRV in advance and further eliminating such losses.
To the best of our knowledge, the most closely related work could be {\eosafe}~\cite{he2021eosafe} and WASAI~\cite{chen2022wasai}, which mainly detect the rollback attack based on symbolic execution and fuzzing, respectively. However, since both frameworks do not consider important state dependencies in smart contracts, they can only detect a sub-type of state-reverting related attacks (i.e., rollback which causes profit-gain). Similarly, eTainter~\cite{ghaleb2022etainter} and Madmax~\cite{grech2018madmax} propose a framework to detect DoS vulnerabilities which are also caused by state-reverting.
Moreover, both {\eosafe}~\cite{he2021eosafe} and WASAI~\cite{chen2022wasai}, are specifically designed for smart contracts in WASM language~\cite{WASM}. Due to the fundamental differences between Solidity contracts and WASM contracts, these frameworks can not be applied to detecting SRVs in Solidity. For example, to detect rollback vulnerability, {\eosafe}~\cite{he2021eosafe} relies on locating specific statement \textit{send\_inline} for invoking an inline action in WASM smart contracts. However, in Solidity there are no such inline actions as well as the \textit{send\_inline} keyword.


\subsection{Scope of Our Work} \system{} is designed to be a generic framework for detecting SRV vulnerabilities in Solidity smart contracts. 
The assumption of \system{} is similar to other smart contract vulnerability detection frameworks, such as SmartDagger~\cite{liao2022smartdagger} and Slither~\cite{feist2019slither}. More specifically, \system{} statically analyzes smart contract bytecode and precisely reports whether a given smart contract contains state-reverting vulnerabilities. 
Since \system{} performs analysis at the bytecode level rather than source code, it is applicable for a number of security vetting scenarios such as large-scale third-party auditing, self-security inspection, etc.
Compared to other research~\cite{ghaleb2022etainter, he2021eosafe, chen2022wasai} which only focuses on specific blockchain platforms (e.g., eTainter for Ethereum and {\eosafe} for EOSIO~\cite{EOSIO}), \system{} can benefit more blockchain ecosystem that based on Solidity smart contracts (e.g., Etherum~\cite{Ethereum}, Tron~\cite{Tron}, and BNB Chain~\cite{BNB}).

\section{Design of \system{}}
\label{sec:overview}

In this section, we present the high-level design of \system{}. We first highlight the key challenges in detecting SRV. Then, we present how \system{} overcomes these challenges in detail.

As mentioned earlier, the root cause of SRVs is that certain state variables can be affected or manipulated by external calls via the state-reverting mechanism. Therefore, a straightforward idea to identify SRVs has two key steps: (1) Identify state dependencies across different state variables and function invocations, and (2) Identify SRVs based on the impacts of such dependencies.

\begin{enumerate}[(1).]
\item \textbf{Identify state dependencies.} Similar to prior research for identifying state-inconsistency bugs~\cite{bose2022sailfish}, we can construct a state dependency graph that represents such dependency relationships across different smart contracts and function calls. Turning to the example shown in Figure ~\ref{fig:SD}, the state variables \textit{SheepToken} and \textit{WolfToken} can be written by the function \textit{MintToken} (line 9, 11) and further accessed by the assert statement (line 14) of the function \textit{Withdraw}. In this case, we say there is a state-dependency relationship between function \textit{MintToken} and \textit{Withdraw} for state variable \textit{SheepToken} and \textit{WolfToken}.

 \item\textbf{Identify SRVs}. For a specific security-sensitive state variable, we can check whether there is a valid path (call chain) along the state-dependency graph, allowing an external call to arbitrarily manipulate it. Again, taking the example in Figure~\ref{fig:SD}, for the state variable \textit{balance} modified in \textit{Withdraw} (line 15), it can be actually affected by the adversary due to the dependency relation between \textit{MintToken} and \textit{Withdraw}. To this end, we identified that the contract \textit{TokenGame} is with a state-reverting vulnerability.
\end{enumerate}

\begin{figure}[t]
\begin{lstlisting}
contract TokenGame {
  mapping(address => uint256) public SheepToken;
  mapping(address => uint256) public WolfToken;
  mapping(address => uint256) public Earning;
  ...
  function MintToken(address account) public {
    uint256 seed= (random () >> 245) % 10; 
    if ( seed != 0) {  
      SheepToken[account]++;} 
    else{
      WolfToken[account]++;} }

  function Withdraw(address account,unit amount) public{
    require(SheepToken[account]>0||WolfToken[account]>0);
    tranferForm(address(this), account, amount); }
    
  function PlaytoEarn(address account,unit tokenId) public{
    if(isWolf(tokenTraits[tokenId])) 
      Earning[account]=Earning[account]*(2-Rate); }
  ...
}
\end{lstlisting}
\caption{The example for vulnerable smart contract with two types of state dependency.}
\label{fig:SD}
\end{figure}

\subsection{Challenges and Solutions}

With the increasing complexity of smart contracts, establishing an effective and complete state-dependency graph (SDG) for SRV detection is by no means trivial. 
Previous research~\cite{bose2022sailfish} performs state-dependency analysis for other purposes (e.g., \textsc{Sailfish} for detecting state inconsistency bugs), these works only cover certain explicit dependencies such as state R\&W relationship and related control flow, which are not sufficient for SRV detection. 
Our work complements and extends previous research by considering two new types of dependencies:  namely, assert-related state dependency (ASD) and temporal-ordered state dependency (TSD), allowing us to detect new vulnerabilities related to state dependency (i.e., SRV).

\para{C1: Extracting assertion-related state dependency (ASD)} ASD refers to dependencies related to statements for access control. Solidity uses reserved keywords such as \texttt{assert, revert, require} for this purpose, and some states can be either explicitly (as function parameters), or implicitly (transaction revert) affected by such statements. If state \textit{S} can be written by function $f_{A}$ and further accessed by assert statement in another function $f_{B}$, we say there is an ASD between function $f_{A}$ and $f_{B}$ for state \textit{S}. In other words, the function $f_{B}$ can execute only when state \textit{S} meets the required condition. Otherwise, function $f_{B}$ fails, and the whole transaction will be reverted. 
Previous work~\cite{liu2021characterizing} showed that 82.28\% of smart contracts take assertion-related statements for access control.
Unlike other explicit dependencies such as state read and write, extracting ASD requires analyzing specific control-flow and data-flow related to assert statements in Solidity.
For example, in Figure~\ref{fig:SD}, for state variables \textit{SheepToken} and \textit{WolfToken}, the state-dependency relationship between \textit{MintToken} and \textit{Withdraw} is an assertion-related state dependency, because function \textit{Withdraw} depends on state variables \textit{SheepToken} and \textit{WolfToken} due to the effect of the assertion statement (i.e., control flow), and these state variables depend on function \textit{MintToken} which writes on it (i.e., data flow).


To overcome this challenge, \system{} performs a fine-grained control-flow and data-flow analysis to recover the program logic by analyzing the smart contract bytecode. In this way, based on the semantics of specific assert statements in Solidity, \system{} establishes the dependencies across different functions for different contract states. 



\begin{figure*}[t]
\centering
\includegraphics[width=6.3in]{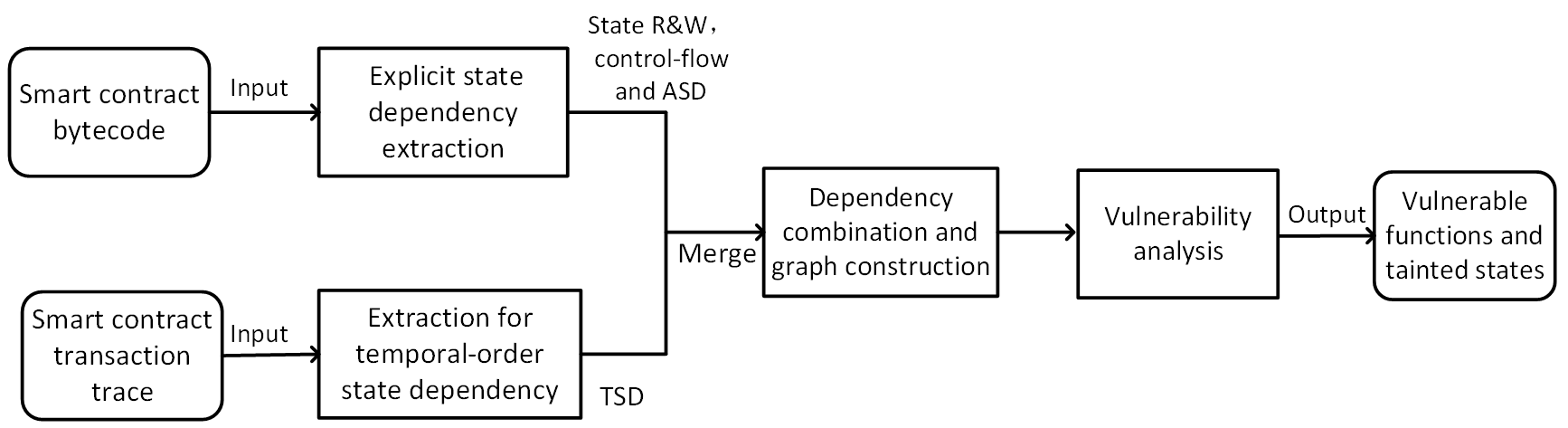}
\caption{The workflow of \system{}.} 
\label{fig:overview}
\end{figure*}

\para{C2: Extracting temporal-ordered state dependency (TSD)} 
As mentioned earlier, smart contracts are executed as transactions on the blockchain. TSD refers to the state dependencies caused by the transaction order of multiple transactions in smart contracts. More specifically, since contract users invoke the contract via a specific transaction sequence, the states produced by contract functions at a certain point could be affected by the temporal order of these transactions. 
For example, as shown in Figure~\ref{fig:SD}, according to the transaction order of this smart contract, we know that the contract user needs to mint the token (i.e., invoking \textit{MintToken}) before playing the game (i.e., invoking \textit{PlaytoEarn}). Therefore, we say there is a temporal order dependency between \textit{MintToken} and \textit{PlaytoEarn}.
The importance of TSD has been highlighted by prior works~\cite{kolluri2019exploiting,so2021smartest,su2022effectively}, as manipulating the transaction order of specific contracts can cause severe vulnerabilities or financial losses.
Unfortunately, the semantics of such temporal order across functions can not be easily recovered at the smart contract bytecode as the transaction order is determined by the application scenarios of a specific smart contract that is related to its functionalities (e.g., a smart contract for gambling). 


To overcome this challenge, \system{} leverages the fact that TSD information is well-preserved on the transaction history of the smart contract, as contract users generally invoke the contract functions according to the specific transaction order. To this end, \system{} leverages a finite state machine to learn the temporal order of smart contract functions from the historical transaction traces, and further extracts the TSD.

\para{C3: SRV identification} The last challenge lies in how to accurately identify SRVs based on the constructed SDG. As mentioned earlier, previous approaches~\cite{grech2018madmax, ghaleb2022etainter} mainly detect SRV-related vulnerabilities based on ad-hoc heuristics such as abnormal gas consumption~\cite{grech2018madmax}. These approaches are not generic and can only cover a sub-type of SRV vulnerabilities (e.g., DoS attacks caused by SRV~\cite{ghaleb2022etainter}).

Different from prior research, \system{} highlights the root cause of SRVs and uses them as SRV indicators to support vulnerability detection in a generic manner. More specifically, \system{} considers the following two key elements as the SRV indicator: (1) Whether a security-sensitive state is non-deterministic, or it shares dependency relation with other non-deterministic states. For example, as discussed earlier in Figure~\ref{illustrateexample}, in the vulnerable function \textit{MintToken}, the number of tokens (e.g., \textit{wolveToken}) can be affected by state variable for storing token values can be affected by the statement that generates a random number.
(2) Whether the sensitive state lacks appropriate access control. Based on the state-dependency graph of a particular state, \system{} detects if necessary access control is missing, for example, checking the origin address of a transaction or checking the balance of a specific token).
Finally, if an external contract can be invoked and reaches states with the above SRV indicators, we consider the analyzed contract vulnerable.

\subsection{Workflow of \system{}}

\system{} takes both the smart contract bytecode and its corresponding transaction traces (i.e., history data) as its input, and finally reports the vulnerability as a set of vulnerability traces. A vulnerability trace contains function calls from the vulnerable function to tainted state variable(s) that can be affected by external call(s). Figure~\ref{fig:overview} shows the workflow of \system{} with the following steps:



\begin{enumerate}[S1.]

\item \textbf{Pre-processing and ASD extraction.} Similar to prior work~\cite{liao2022smartdagger}, \system{} first recovers the control flow and data flow of the whole smart contract as the pre-processing step. Then, it identifies ASD from the program logic of smart contracts.

\item \textbf{Temporal-ordered state dependency
extraction.} In the second step, \system{} analyzes historical transaction traces and extracts the TSD.

\item \textbf{Dependency combination and graph construction.} Then, \system{} merges four types of dependency, i.e., the state R\&W dependency, control flow, ASD, and TSD, to generate the state dependency graph (SDG) for SRV detection.

\item \textbf{Vulnerability detection.} Lastly, based on the constructed SDG, \system{} identifies all the SRV indicators and finds out vulnerability traces with vulnerability-specific rules.





\end{enumerate}


\section{Approach Details}
\label{sec:methodology}

Now we elaborate on the details of each step in \system{}. 

\subsection{Pre-processing and ASD Extraction}
\label{sec:explicit}

\para{Pre-processing} \system{} utilizes 
\textit{SmartDagger}~\cite{liao2022smartdagger}, a state-of-the-art static analysis tool to recover the control flow and data flow from the bytecode of a given smart contract. Since SmartDagger is designed for detecting cross-contract vulnerabilities, it can construct a more complete control flow and data flow for cross-functions (contracts) invocations, compared to other similar tools (e.g., Mythril~\cite{mythril}, Slither~\cite{feist2019slither}). Specifically, \textit{SmartDagger} decompiles the bytecode of the smart contract to generate the intermediate representation (IR). Then, based on the IR and function invocation information, it outputs the constructed the control-flow and data-flow graph.

\para{ASD extraction} After recovering the program logic, \system{} identifies the ASD from the extracted control flow and data flow. Here, \system{} first leverages a similar approach as in \textsc{Sailfish}~\cite{bose2022sailfish} to extract the basic read and write (R\&W) dependency for state variables. In addition, \system{} extracts ASD with the following heuristics: 

\begin{itemize}
    
    \item A function $M_r$ reads the state variable $S_d$ as a condition within the assertion statement (i.e., \texttt{revert, assert, require});
    
    \item There is another function $M_w$ which writes on the same state variable $S_d$. 

\end{itemize}

If the above two conditions are satisfied, \system{} determines that function $M_r$ has an ASD on function $M_s$. To this end, it adds a directed edge $e_{r}(M_r, M_w)$ to denote this dependency. Finally, all ASD edges of the smart contract can be denoted as a set of directed edges $E_r= \left\{ e_{r}(M_r, M_w)| M_r, M_w \in M \right\}$.

\begin{figure}[tbp]
\centering
\includegraphics[width=3.4in]{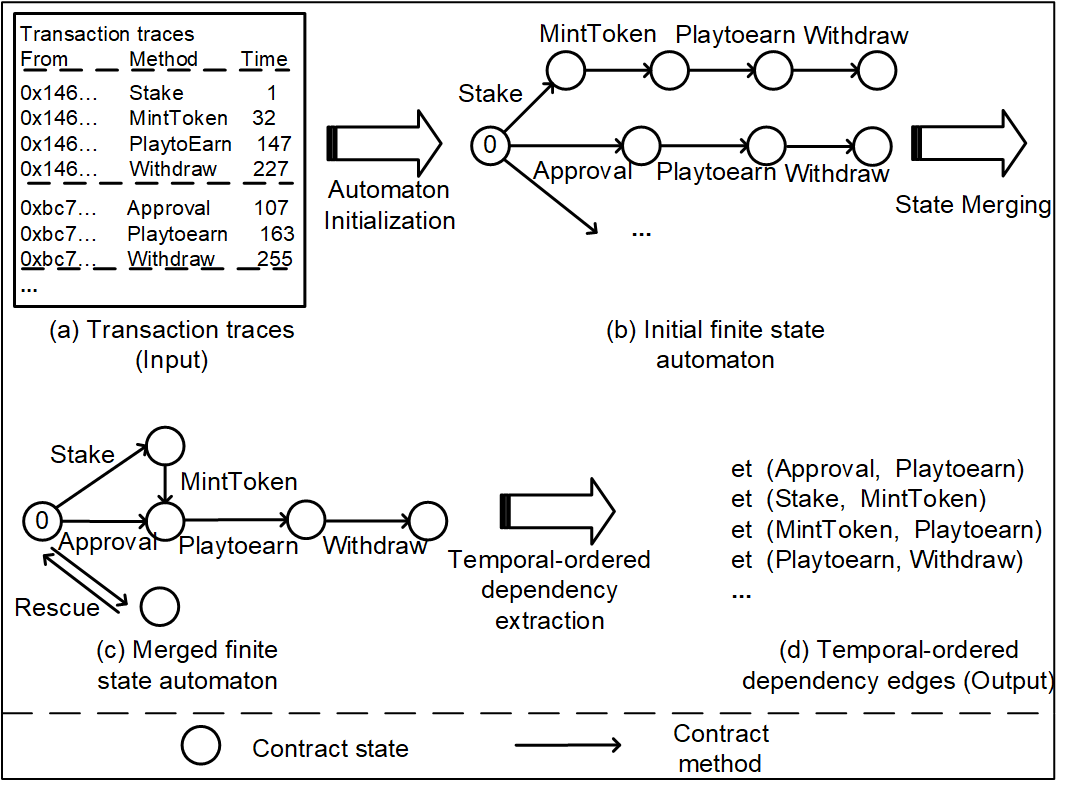}
\caption{Details of the finite state machine.} 
\label{fig:timedautomata}
\end{figure}
\subsection{TSD Extraction}
\label{subsec:model-extraction}

\system{} extracts the TSD from the historical transaction traces of smart contracts. Take the transaction traces in Figure~\ref{fig:timedautomata} (a) as an example. The user ``0x146...'' executes a group of transactions in line with the specific order (i.e., \textit{MintToken}, \textit{PlaytoEarn}). Assuming that such transaction information is sufficiently effective, we can identify that \textit{PlaytoEarn} has a TSD on \textit{MintToken}. Obviously, practical transaction order is more complex. It is a demand for an effective approach to learning about the TSD from transaction traces. 

\system{} leverages a finite state machine (FSM)~\cite{Leonardo2017GKtail} to extract TSD. A finite state machine is a mathematical model which can accurately describe the state transitions of specific subjects \cite{Pastore2022TKT}. In our case, FSM is used to represent transitions of contract states during the execution of transactions. More specifically, each node represents a specific state of the smart contract with a set of state variables, each edge refers to a specific state transition caused by a transaction execution, and the value of each edge is the function invoked by the transaction. We denote this FSM as the following tuple $(S, s_{0}, Tr, M, T)$ where

\begin{itemize}

\item $S$ represents the set of contract states.
\item $s_{0} \in S$ represents the initial state of the smart contract.
\item $T_r$ represents the set of transactions in the smart contract.
\item $M$ represents the set of values that correspond to the functions invoked by the transactions.
\item $T$ represents the transition relation $T: S\times T_r \rightarrow S$

\end{itemize}

 Note that each state transition is caused by the transaction that executes on the predecessor state and finally reaches the successor state, denoted as $T: S\times Tr \rightarrow S$. 
 


%

\para{FSM construction} Below we introduce how the FSM works for TSD extraction. The input of FSM is a set of transaction traces. Specifically, each trace represents the transaction sequence that is invoked by a specific contract account, and each transaction contains information about the address of the user, invoking function, and invoking clock. 
The output of FSM is TSD edges, also denoted as a set of directed edges $E_t= \left\{ e_{t}(M_{t1}, M_{t2}) | M_{t1}, M_{t2} \in M \right\}$. For instance,  Figure~\ref{fig:timedautomata} (a) and (d) show an example of the transaction traces, and the finally extracted TSD edges, respectively. 

The FSM generates such output through the following steps:
\begin{enumerate}[(1).]
\item \textbf{Initialization.} As shown in Figure~\ref{fig:timedautomata} (b), \system{} first combines all transaction traces into a tree-shaped machine. In the machine, each trace corresponds to a single branch of the tree. Besides, the machine is labeled with relative functions that correspond to the transactions in the traces.

\item \textbf{State merging.} Then, \system{} merges the states of the initial FSM via analyzing the transaction equivalence and transaction subsumption of different states. In this way, the FSM can be simplified into a sufficiently compact model. For instance, Figure~\ref{fig:timedautomata} (b) and (c) show the FSM update before and after the merging process.
\item \textbf{TSD extraction.} Finally, \system{} extracts the TSD from the merged FSM. Specifically, \system{} determines the TSD according to the temporal order of state transition in the FSM. For example, the temporal-ordered state dependency edges in Figure~\ref{fig:timedautomata} (d).



\end{enumerate}


\begin{figure}[tbp]
\includegraphics[width=3.3in]{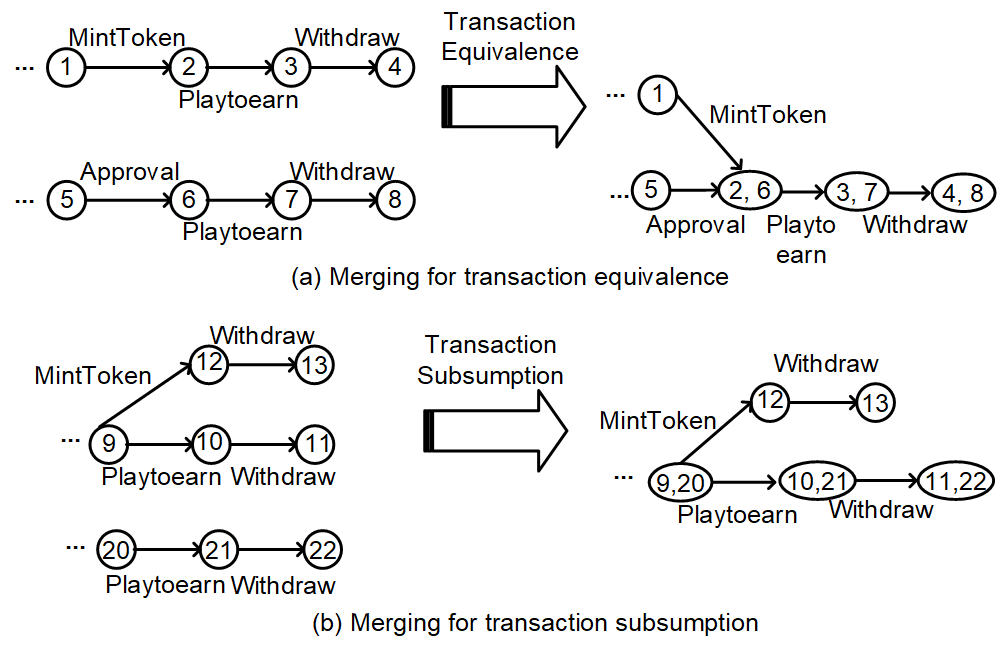}
\caption{Examples for illustrating transaction equivalence and transaction subsumption.} 
\label{fig:transaction equivalence and transaction subsumption}
\end{figure}


\para{State Merging} 
\system{} merges all the equivalent states and subsumed states in the initial FSM to shrink its size. This process is done in a way similar to methods proposed by prior works~\cite{Pastore2022TKT, Leonardo2017GKtail}. We take two additional examples shown in Figure~\ref{fig:transaction equivalence and transaction subsumption} to illustrate this process. As can be seen from part (a), transactions $s_{2}$ and $s_{6}$ are equivalent because the share the same future functions (i.e,  $\left\{ Playtoearn, Withdraw\right\}$). Therefore, transactions [$s_{2}$, $s_{6}$], [$s_{3}$, $s_{7}$], and [$s_{4}$, $s_{8}$] are merged. Similarly, the process of transaction subsumption is presented in Figure~\ref{fig:transaction equivalence and transaction subsumption}(b).

\subsection{Dependency Graph Construction}
\label{subsec:WSDG}

\system{} merges the previously extracted control flow, state R\&W dependency, ASD, and TSD together to construct the fine-grained state-dependency graph (SDG), to facilitate SRV detection in a more effective manner. Below we give more details about the SDG and its construction process.

\para{Data structure} The SDG constructed by \system{} is denoted as a triple $G_{s}=(N_{s},E_{s},X)$. Specifically, SDG encodes the following information: (1) The nodes in SDG are a set of state variables and basic block nodes representing program operations. Here, $S$ denotes the set of state variables, $B$ denotes the set of basic block nodes. We include basic blocks as SDG nodes because basic blocks provides important dependency information related to state variables (i.e., function invocation). Therefore, we say $N_{s}:=\left\{ S\cup B\right\}$. (2) The edges in SDG are a set of control-flow edges, state R\&W dependency edges, ASD edges, and TSD edges. $X(E_{s}) \rightarrow \left\{C, R\&W, ASD, TSD \right\}$ is a labeling function that maps an edge to one of the four types. 



\begin{figure}
\centering
	\includegraphics[width=3.3in]{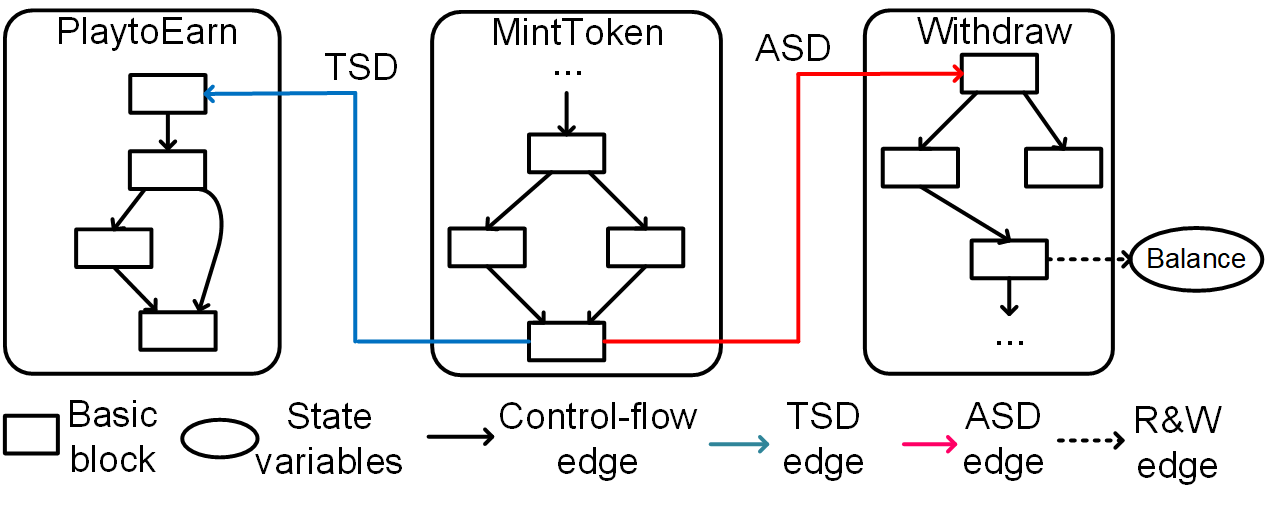}
    \caption{The process of fine-grained SDG construction for the running example described in Figure~\ref{fig:SD}.} 
    \label{fig:WSDG}
    
\end{figure}

\para{SDG construction} \system{} constructs SDG by incrementally adding ASD edges, TSD edges, and state R\&W dependency edges to the control flow graph. For each ASD, \system{} searches the control-flow graph and finds out the source and target basic blocks, and connects them with a directed edge. Note that if function A has an ASD on function B, the source of such ASD is the end site (i.e., the last statement such as \textit{return}) of function B, and the target of ASD is the start cite of function A. After that, \system{} adds the directed edge between the source and target for each ASD. Similarly, \system{} adds the TSD edges to the graph in the same manner. For state R\&W dependency edges, \system{} searches for basic block nodes that read or write the state variables and adds the directed edge between these basic blocks and state variables.
Figure~\ref{fig:WSDG} shows an example of the constructed SDG. As can be seen, the ASD edges and TSD edges are labeled with different colors.

\subsection{Vulnerability Detection}
\label{vulnerabilitydetection}

\para{SRV indicator} Unlike prior works that detect SRV by identifying the side effect of a specific sub-scenario (e.g., identifying gas sufficiency), \system{} takes a set of generic rules for locating vulnerable functions according to the root cause of SRV. More specifically, for each function in smart contracts, \system{} identify it as an SRV indicator (i.e., vulnerable function) only if (1) the state variables can produce uncertainty or have state dependency with other uncertainty states and (2) along the state dependency path, there is lack of correct access control. We formalize the generic rules as vulnerable indicator as follows.

\vspace{4pt}
\noindent\framebox[235pt][r]{$isUncertainty (var_{state}) \wedge isLackOf(C_{acc})$ \ for each $f$ \ \ \  (1)} \\ 

In this formula, $f$ is the function in the smart contract, $var_{state}$ represents the state variable in the function $f$, and $C_{acc}$ refers to the correct access control condition. Particularly, the former condition is scenario-specific, and the latter condition is generic to all scenarios of SRV. Further, based on the above formula, table~\ref{table-idicatoring} summarizes the detailed rules for analyzing two sub-scenarios of SRV covered by \system{}.

\begin{table}[h]
\small
\caption{Vulnerability indicator rules for locating two sub-scenarios of SRV}
\begin{tabular}{|l|c|}
\hline
Vulnerability type               & Vulnerability indicator rule                                                                                                           \\ \hline
R1-Profit-gain attack             & \begin{tabular}[c]{@{}c@{}}$ isRandomness (var_{state}) \wedge  isLackOf(C_{acc})  $  \end{tabular}                                                                      \\ \hline
R2-DoS attack             & \begin{tabular}[c]{@{}c@{}}$(isInLoop (external call)$ \\ $\wedge isModified (var_{state})) \wedge isLackOf(C_{acc})$\end{tabular} \\ \hline

\end{tabular}
\label{table-idicatoring}
\end{table}

With the above-illustrated SRV indicator, \system{} takes three key steps to detect SRV. 

\begin{table*}[t]
\small
\caption{EVM instructions defined as taint sources and taint sinks by \system.}

\begin{tabular}{|l|l|l|}
\hline
                        & Type                                     & EVM Instruction or Keyword or Statement                             \\ \hline
\multirow{2}{*}{Source} & (1) Parameter passed by contract invoker & CALLDATALOAD, CALLDATACOPY, CALLER, ORIGIN, CALLVALUE, CALLDATASIZE \\ \cline{2-3} 
                        & (2) Parameter of public function         & Public, External                                                    \\ \hline
\multirow{2}{*}{Sink}   & (1) External calls                       & CALL, CALLCODE, STATICCALL, DELEGATECALL                            \\ \cline{2-3} 
                        & (2) State variables                      & SSTORE, BALANCE, ADDRESS, SRV indicators                            \\ \hline
\end{tabular}


\label{table.taint}
\end{table*}

\textbf{Step-1: Finding SRV indicator(s).} Firstly, \system{} screens for the SRV indicators on the SDG based on our proposed generic rules. This process is modeled as a process of graph query to locate the vulnerable function from SDG. For example, for a profit-gain attack, \system{} inspects whether there exist statements that randomly modify a balance variable (e.g., whether statements with the EVM instruction \texttt{SSTORE} is in an execution lock depending on randomness) as well as the lack of access control for forbidding arbitrary external calls.

\textbf{Step-2: Finding the entry trace of external contract.} After identifying the SRV indicators, \system{} searches for the entry trace for each SRV indicator and utilizes taint analysis to inspect whether it can be accessed by an external attacker. To model this execution flow, \system{}  makes the taint propagate from the entry point (e.g., public function) of the contract, and observes whether the taint can propagate to the vulnerability indicators.

\textbf{Step-3: Finding tainted state variable(s) affected by state-reverting.} After finding the entry trace, \system{} continues to perform the forward taint propagation on the SDG and computes the tainted state variables. Apparently, these state variables can be affected by the subverted flow of the vulnerability. Finally, all the tainted functions and state variables are reported as the vulnerable trace that describes the \bugname{}.

To perform taint propagation, the taint sources can be divide into two types: the parameters passed by contract callers and parameters of public functions. The taint sinks of \system{} consist of either those external calls, or state variables of the smart contracts (including the SRV indicators). More detailed information regarding the taint sources and sinks are summarized in Table~\ref{table.taint}.

\begin{figure}[h]
\centering
	\includegraphics[width=3.5in]{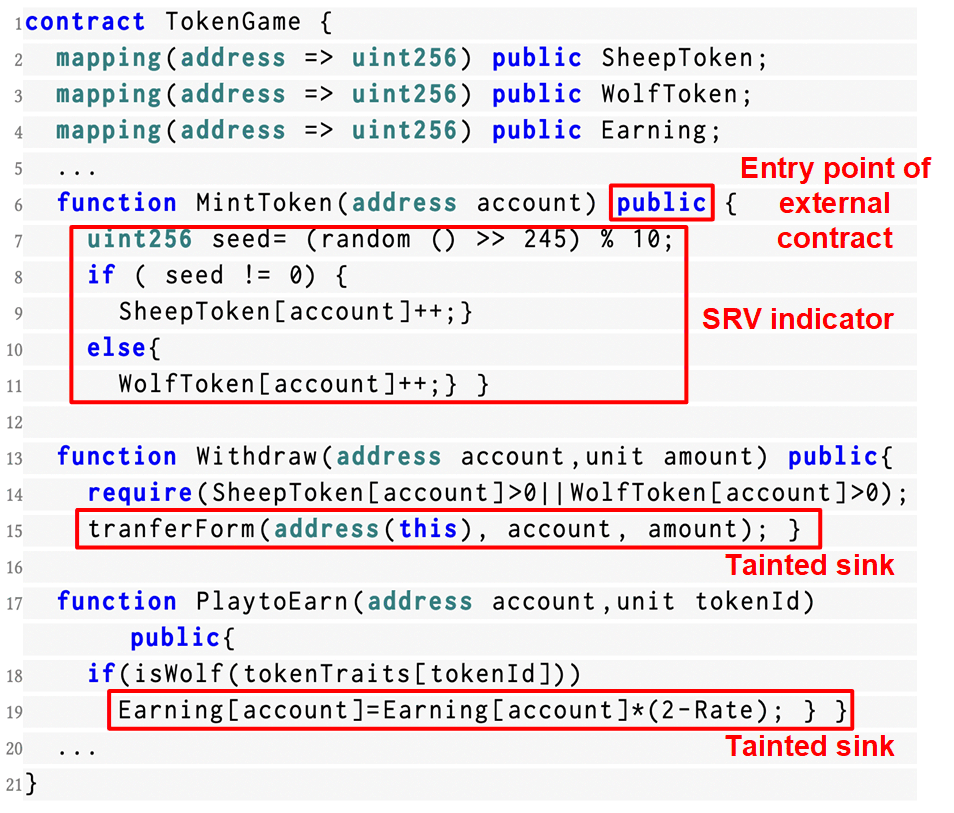}
    \caption{The process of vulnerability detection for the motivating example.} 
    \label{detection}
    
\end{figure}

Again, we take the running example shown in Figure~\ref{detection} as an instance to show this process. 
For this smart contract, \system{} searches for the SRV indicators on the SDG. \system{} identifies that the function \textit{MintToken} satisfies Step-1 because \textit{MintToken} utilizes a random number (seed) which determines the modification on state variable (i.e., SheepToken and Wolfoken) without access control. Then, \system{} performs the Step-2, and identifies that \textit{MintToken} is a public function that can be accessed by an external attacker. Lastly, \system{} performs the Step-3 by screening for state variables tainted by the vulnerability through data flow analysis.
As a result, \system{} reports the vulnerability traces in contract \textit{TokenGame} as (1) \textit{MintToken} $\rightarrow$ $\left\{SheepToken,WolfToken\right\}$, (2) \textit{MintToken} $\rightarrow$ \textit{Playtoearn} $\rightarrow$ $\left\{Earning\right\}$, (3) \textit{MintToken} $\rightarrow$ \textit{Withdraw} $\rightarrow$ $\left\{Balance\right\}$.

\section{Evaluation}
\label{sec:eval}

In this section, we first present our experimental setup and the two datasets for evaluation (the manual-labeled SRV dataset and the large-scale smart contracts dataset). Then, we show the evaluation results of \system{} in terms of false positives and false negatives over the manual-labeled SRV dataset. Lastly, we discuss the results of our large-scale analysis and identified new SRVs in the wild.

\subsection{Implementation and Evaluation Setup} 

We implement \system{} with around 3,400 Line-of-Code in Python 3.8.10. All the experiments in our evaluation are conducted on an Ubuntu 20.04 server that is equipped with one Intel i9-10980XE CPU (3.0GHz), one RTX3090 GPU, and 250 GB RAM.

\para{Dataset and ground-truth establishment} We utilize the following two datasets for our evaluation experiment.

\textit{Manual-labeled SRV Dataset ($D_{srv}$)}. This dataset constructs the ground truth for evaluating the effectiveness of \system{}. More specifically, we manually collected and annotated a total number of 91 SRVs from 47 smart contracts\footnote{each smart contract has a unique address in Ethereum}. Particularly, 27 SRVs belong to \textit{profit-gain} and the other 64 SRVs belong to \textit{DoS}.
Among this dataset, 17 SRVs (from 11 smart contracts) are collected based on public-reported attack incidents. 
To the best of our knowledge, this is the most comprehensive collection of SRVs from the public sources. We manually locate the related SRVs by reviewing such reports.
The rest 74 SRVs (from 36 smart contracts) are manually collected by inspecting financial-related contracts (e.g., Wallet and Gambling) from popular DApps.
%
To avoid bias, three researchers participated in the annotation process. Every researcher separately performs the annotation. Only the vulnerability agreed by all three researchers are confirmed as a valid SRV. 
We selectively explored financial applications for building $D_{srv}$, as their SRVs may cause severe impact (e.g., direct money losses). However, this does not mean \system{} is specifically tailored or biased to financial applications, as the patterns of SRVs are generic to all contracts.

\textit{Large-scale Dataset ($D_{large}$)}. To show the effectiveness of \system{} in finding SRVs in the real world, we utilize the second dataset, which contains 47,351 smart contracts. The dataset is an open dataset that is proposed in the well-known empirical study~\cite{durieux2020empirical}.

\para{Evaluation Metrics} We summarize the following research questions (RQs) to evaluate \system{}.

\begin{enumerate}[RQ1.]
	\item How effective is \system{} in terms of detecting the state-reverting vulnerability? 
 
	\item For state dependency analysis, how do the extracted ASD and TSD help for detecting \bugname{}? 
    
    \item How does \system{} perform compared to other state-of-the-art mechanisms in terms of detecting SRV?
    
	\item Can \system{} detect \bugname{} from real-world smart contracts?  
\end{enumerate}

\begin{table}[htbp]

\caption{Overall effectiveness for \system{} on the Manual-labelled SRV Dataset ($D_{srv}$).}

\begin{tabular}{l|ccc|ccc}
\hline
Attack exploits SRV                                  & \multicolumn{3}{c|}{Precision}                                               & \multicolumn{3}{c}{Recall}                               \\ \hline
                                        & \multicolumn{1}{r}{TP} & \multicolumn{1}{r}{FP} & \multicolumn{1}{r|}{rate} & TP & \multicolumn{1}{r}{FN} & \multicolumn{1}{r}{rate} \\ \hline
\multicolumn{1}{l|}{Profit-gain attack} & 24                       &5                        & 82.76\%                           & 24 & 4                      & 85.71\%                    \\
DoS attack                              & 58                       & 7                     & 89.23\%                           & 58 & 6                      & 90.63\%                    \\ \hline
Total                                   & 82                       & 12                       & 87.23\%                           & 82 & 10                     & 89.13\%                    \\ \hline
\end{tabular}

\label{table.rollback}
\end{table}
\begin{table*}[h]

\caption{Comparison results between \system{} and the other two baselines over the Manual-labeled SRV Dataset ($D_{srv}$).}

\begin{tabular}{l|rrr|rrr|rrr}
\hline
Approach           & \multicolumn{3}{c|}{\begin{tabular}[c]{@{}l@{}}\system{} w/o ASD and TSD\end{tabular}}                                               & \multicolumn{3}{c|}{\begin{tabular}[c]{@{}l@{}}\system{} w/o TSD\end{tabular}}                                               & \multicolumn{3}{c}{\system{}} \\ \hline
                   & \multicolumn{1}{r}{TP} & \multicolumn{1}{r}{FN} & \multicolumn{1}{r|}{recall} & \multicolumn{1}{r}{TP} & \multicolumn{1}{r}{FN} & \multicolumn{1}{r|}{recall} & TP    & FN    & recall     \\ \hline
Profit-gain attack & 12                     & 16                     & 42.86\%                     & 18                       & 10                       & 64.28\%                            & 24    & 4     & 85.71\%    \\
Dos attack         & 42                     & 22                     & 65.63\%                     & 53                       & 11                       & 82.81\%                            & 58    & 6     & 90.63\%    \\ \hline
Total              & 54                     & 38                     & 58.70\%                     & 71                       & 21                       & 77.17\%                            & 82    & 10    & 89.13\%    \\ \hline
\end{tabular}
\label{table:cross-method dependency}
\end{table*}

\subsection{Effectiveness of \system{}}


To answer RQ1, we ran \system{} on the manual-labeled SRV Dataset ($D_{srv}$) to evaluate its precision and recall. For example, we give the same time budget (i.e., 10-mins timeout) for every smart contract in the dataset. The precision (false positive) and recall (false negatives) are computed by manually inspecting the reported results and comparing these results with the ground-truth data of $D_{srv}$ (i.e., 91 vulnerabilities in the 47 contracts).

Table~\ref{table.rollback} presents the overall results. As can be seen, \system{} achieves a high recall (i.e., 89.13\%) and precision (i.e., 87.23\%). 


\para{False positives and false negatives} We manually inspect the false positives and false negatives introduced by \system{}. Among 12 false positives, our further inspection shows that most of them are caused by the limitation of control flow analysis produced by SmartDagger (i.e., the existing analyzer used in \system{}). For example, \system{} reports the false results of vulnerable functions due to the fact that SmartDagger fails to locate the function borders. To address these false positives, \system{} can be improved by integrating a more advanced bytecode analyzer for recovering function boarders. For the 10 false negatives, most of them are because they rely on third-party data which is not controlled by the blockchain platform (e.g., a pseudo-random number that relies on oracle). In fact, the problem can not be addressed by any static analysis approach like ours, as they require third-party data outside the blockchain.


\subsection{Effectiveness of ASD and TSD}

To answer RQ2, we evaluate the effectiveness of individual components in \system{}, i.e., ASD and TSD. As mentioned earlier in Section~\ref{sec:explicit} and~\ref{subsec:model-extraction}, ASD and TSD are the two important advantages possessed by \system{},
which ensure the soundness of vulnerability analysis (i.e., avoiding false negatives). For instance, \system{} can utilize such advantages to perform more taint tracking so that it can identify more vulnerability traces, in contrast to those approaches without considering ASD and TSD. Hence, the effectiveness of ASD and TSD is reflected on the recall rate. We compare the current design with two baseline approaches. More specifically, \system{} without considering ASD\&TSD, as well as \system{} without considering TSD. We ran these baseline frameworks on $D_{srv}$.

\para{Compare to \system{} without TSD} 
We evaluate the effectiveness of TSD by comparing \system{} to the baseline approach without considering TSD.
The column 2 and 3 in table~\ref{table:cross-method dependency} show the recall rate for this comparison. Due to ignoring the TSD, the total recall rate of the baseline approach without considering TSD is only 77.17\%, the recall rates of  both \textit{Profit-gain attack} and\textit{Dos attack} drop obviously. Particularly, the recall rates of \textit{Profit-gain attack} drop more rapidly than that of \textit{DoS attack}. In contrast, \system{} maintains a good performance (over 85\%) in terms of two types of scenarios. 
In conclusion, the extracted TSD effectively helps \system{} improve the recall for SRV detection. Particularly, the extracted TSD is more important for the detection of \textit{Profit-gain attack} than \textit{DoS attack}.

\para{Compare to \system{} without TSD and ASD} After that, we further evaluate the effectiveness of ASD by performing the comparison between \system{} and the other baseline approach without considering ASD and TSD. The columns 1 and 3 in table~\ref{table:cross-method dependency} shows the recall rate for this comparison. Due to ignoring ASD and TSD, the total recall rate of this baseline approach is only 58.70\%, the recall rates of \textit{Profit-gain attack} and \textit {DoS attack} drop more rapidly. And this baseline approach produces more false negative (i.e., a total of 17 new false negatives).  To sum up,  the extracted ASD also helps \system{} improve the recall for SRV detection significantly.

Further, we manually inspected every false negative reported by two baseline approaches. The inspected results show that 17 (i.e., 44.74\%) of the 38 false negative results can be avoided by analyzing ASD, and 11 (i.e., 28.95\%) of the 38 false negative results can be avoided by analyzing the TSD, which are missed by these baseline approaches. For instance,
Figure~\ref{bs1example} shows an example of false negatives, which can be avoided by \system{} with the help of analyzing ASD. For this case, the baseline without ASD and TSD can only report that function \textit{redeem} contains a DoS attack vulnerability, as it cannot identify the ASD between functions \textit{redeem} and \textit{transferFrom}. Nevertheless, if we analyze such state dependency, we can find that function \textit{redeem} is a vulnerability indicator, and it can taint the function \textit{transferFrom}.   \system{} avoids such false negatives, as it extracts the ASD edge and consequently finds all the vulnerabilities alongside it.

\subsection{Effectiveness of Vulnerability Indicator Analysis}

As mentioned earlier in Section~\ref{vulnerabilitydetection}, another advantage processed by \system{} is the vulnerability indicator analysis in the detection, which helps for improving the precision of \bugname{} identification. The effectiveness of vulnerability indicator analysis is reflected in the precision rate. To evaluate the effectiveness of vulnerability indicator analysis,
we compare the precision rate of \system{} with state-of-the-art tools (i.e. eTainter~\cite{ghaleb2022etainter} Madmax~\cite{grech2018madmax} and Slither~\cite{feist2019slither}).
As neither of the three prior tools can support identifying \textit{Profit-gain attack} vulnerability, we evaluate the effectiveness of vulnerability indicator analysis by comparing \system{} with three prior tools for identifying \textit{DoS attack} vulnerability. We ran all of these tools on the large-scale dataset ($D_{large}$) to evaluate their precision.

As shown in table~\ref{table.dos}, \system{} presents much higher precision (i.e., 84.16\%) than the three prior tools. Our further investigation finds that most of the false positives introduced by prior tools can be avoided through our proposed vulnerability indicator analysis. In Figure\ref{VIexample}, we present an example to show how \system{} avoids the false positive, which eTainter, Madmax, and Slither report. All the prior tools mistakenly report that function \textit{refund} contains SRV, as function \textit{refund} invoke an external call in a loop (line 7-8). Actually, if we check the access control (line 5), we can find that the function can be only accessed by externally owned accounts rather than contract accounts, which causes function \textit{refund} to be safe. Due to the incomplete detection approaches, prior tools do not identify the access control of the function and consequently report false positive results. In contrast, \system{} can avoid such false positives because it performs multiple checks on function \textit{refund} through the vulnerability indicator analysis and infers that such function has access control and is consequently safe.

\begin{table}[hb]

\caption{Precision rate for different tools on the large-scale dataset ($D_{large}$).}

\begin{tabular}{l|rrr}
\hline
Tool & \multicolumn{3}{c}{Precision}                                                   \\ \hline
              & TP                      & FP                      & rate                         \\ \hline
eTainter       & \multicolumn{1}{c}{178} & \multicolumn{1}{r}{215} & \multicolumn{1}{c}{45.29\%} \\
Madmax        & 19                      & 54                      & 26.03\%                      \\
Slither       & 151                     & 2,555                    & 7.08\%                       \\ \hline
\system{}        & 574                     & 108                     & 84.16\%                      \\ \hline
\end{tabular}

\label{table.dos}
\end{table}

\subsection{Large-scale Analysis for Finding SRVs}

To answer RQ4, we evaluate the performance of \system{} by running \system{} on the large-scale dataset ($D_{large}$) for SRV detection. 
Our domain experts manually inspect these results via majority voting and confirm that \system{} successfully identifies 406 new SRVs from 47,351 real-world smart contracts. Specifically, \system{} reports 771 warnings (including 651 TPs and 120 FPs confirmed manually). 245 of 651 SRVs can be detected by SOTA tools (i.e., Madmax~\cite{grech2018madmax}, Slither~\cite{feist2019slither} and eTainter~\cite{ghaleb2022etainter}). Therefore, SmartState reports 406 (651-245) new SRVs. 
Particularly, we rank the large-scale dataset ($D_{large}$) according to their number of transactions and intercept the smart contracts of the top 2000 transaction number as the ``popular smart contracts''.
We found that 11 SRVs exist in the popular smart contracts. Further, these 11 \bugname{} affects a total asset of 428,600 USD as of manuscript submission. 
Below we discuss two case studies for illustration. 

\begin{figure}[t]
\begin{lstlisting}
contract BsktToken {
  //DoS indicator 
  function redeem (uint256 baseUnits) external { 
    for (uint256 i = 0; i < tokens.length; i++) {
      uint256 amount = baseUnits;
      //state variable tainted by DoS
      require(erc20.transfer(msg.sender, amount)); }}
  //ASD propagate
  function transferFrom(address _from, address _to, uint256 _value) public {
    require(_value <= balances[_from]);
    balances[_from] = balances[_from].sub(_value); 
    //state variable tainted by DoS (cross-method)
    balances[_to] = balances[_to].add(_value);}
    
\end{lstlisting}

\caption{An example of false negative reported by \system{} w/o ASD and TSD, but effectively eliminated by \system{}}
\label{bs1example}
\end{figure}

\begin{figure}[t]
\begin{lstlisting}
contract Lotto {
  address[] internal playerPool;
  function refund() public payable {
    // access control, only EOA account can access
    require(tx.origin = msg.sender)
    require(playerPool.length > 0);
    for (uint i = 0; i < playerPool.length; i++) {
      playerPool[i].transfer(100 finney);} 
    playerPool.length = 0; }
\end{lstlisting}

\caption{An example of false positive reported by eTainter and Madmax, which is avoided by \system{}.}
\label{VIexample}
\end{figure}

\para{Case study 1} at \textit{0xEB834ae72B30866af20a6ce5440Fa598BfAd3a42}. 
The smart contract is from the GameFi \textit{WolfGame}, which ranks 83 in the top 1000 market-value GameFi. Unfortunately, this smart contract contains a SRV. The vulnerability indicator of such SRV is the function \textit{claimSheepFromBarn}, as it leverages a random number to determine the punishment result of free of tax behavior but without forbidding the invocation of an external contract. 
Hence, the function \textit{claimSheepFromBarn} can be attacked by malicious players via invoking the specific call chain (i.e., from \textit{claimManyFromBarnAndPack} to \textit{claimSheepFromBarn}).
For this case, \system{} effectively identifies the vulnerable function and utilizes the state dependency between function \textit{claimSheepFromBarn} and function \textit{mint} to determine that 
state variable \textit{balance} is manipulated by the SRV.

\para{Case study 2} at \textit{0x222222de1914c2b303504e3b035cf46b11fcfc6c}. As of manuscript submission, this smart contract possesses an asset of 157.41 ETH (i.e., 248,396 USD)  and involves 15,465 transactions. Unfortunately, the smart contract contains the SRV. In this case, the vulnerability indicator is the function \textit{pay\_royaltie} because it invokes an external transfer in a loop but misses forbidding the invocation of an external contract. If one of the users refuses the money and makes the transaction fail through an assertion statement, the whole loop fails and locks all the rewards in the contract. For this case, \system{} effectively locates the vulnerability indicator and reports that state variable \textit{withdrawl} and \textit{balance} is tainted by the SRV.

\subsection{Discussion and Limitation}

To sum up, \system{} demonstrates its advantages for detecting \bugname{}. (1) As shown in the evaluation, \system{} can effectively identify \bugname{} for smart contracts, which a few prior works can support. (2) \system{} can precisely find out the root cause (i.e., indicator) of SRV and identify the state variables tainted by SRV by analyzing the state dependency of smart contracts. Hence, \system{} can locate the vulnerability more precisely, in contrast to those approaches devoted to finding out the side effect (i.e., Gas insufficiency) of vulnerability, so it is sufficiently effective and practical. All of the developers, participants, and third-party authorities can leverage \system{} to inspect the security of smart contracts.

\system{} can be further improved in the following aspects: (1) At present, \system{} can support identifying two major types of \bugname{}, and we can further widen its detection scopes so that it can support more newly emerging types of vulnerabilities. (2) To further improve its effectiveness, \system{} can further utilize more precise and robust tools instead of SmartDagger for its program logic recovery analysis.



Below we analyze the soundness and completeness of every design employed in \system{}. Firstly, the explicit dependency construction is limited by program logic recovery (i.e., provided by SmartDagger), resulting in imprecise information and incompleteness.
Secondly, the TSD extraction is limited by the diversity of transaction history data, as finite state machine cannot extract all the TSD, it may introduce unsoundness. Thirdly, SDG construction and vulnerability analysis are complete and sound, as it neither introduces false information nor misses valid information. 


\section{Related Work}

\para{Vulnerability detection for smart contract}
At present, many program analysis tools have been proposed for detecting vulnerabilities in smart contracts. Similar to traditional program analysis, these tools can be divided into static analysis tools and dynamic analysis tools. Specifically, the static analysis tools include Oyente~\cite{luu2016making}, Osiris~\cite{torres2018osiris}, Zeus~\cite{kalra2018zeus}, \textsc{Sailfish}~\cite{bose2022sailfish}, SmartDagger~\cite{liao2022smartdagger}, {\eosafe}~\cite{he2021eosafe}, Ethainter~\cite{brent2020ethainter}, Clairvoyance~\cite{xue2020cross}, MadMax \cite{grech2018madmax}, Manian\cite{nikolic2018finding}, Securify~\cite{tsankov2018securify} and so on. Other tools such as ContractFuzzer~\cite{jiang2018contractfuzzer}, Sereum~\cite{rodler2018sereum}, EOSFuzzer \cite{huang2020eosfuzzer}, Echidna~\cite{grieco2020echidna}, echidna-parade~\cite{groce2021echidna}, sFuzz~\cite{nguyen2020sfuzz}, TXSPECTOR ~\cite{zhang2020txspector}, SMARTIAN~\cite{choi2021smartian}, WASAI~\cite{chen2022wasai} and RLF~\cite{su2022effectively} are based on dynamic testing or analysis. Nonetheless, all these tools are insufficiently effective in detecting SRVs as they do not consider the fine-grained state dependency for vulnerability detection. When the recent work (i.e., \textsc{Sailfish}~\cite{bose2022sailfish}) focuses on detecting state-inconsistency via analyzing the state dependency, the SDG constructed by SailFish can only cover a subset of state dependency (i.e., control flow and state R\&W dependency), which cannot cover the state dependency encountered by SRV detection.

\para{State-reverting vulnerabilities}
There have been limited works closely related to SRVs.
{\eosafe}~\cite{he2021eosafe} and WASAI~\cite{chen2022wasai} focus on detecting the rollback attack (i.e., rollback which causes profit gain)) for smart contract written in WASM language. However, since both of the two frameworks mainly identify the vulnerability via analyzing the heuristic pattern without considering important state dependencies in smart contracts, they can only cover part of the SRV (i.e., profit-gain). Also, they are not sufficiently effective in detecting this sub-type of state-reverting related attacks.
While the two key steps (in Section~\ref{sec:overview}) used for detecting SRVs in \system{} are based on heuristics, our proposed SRV indicators are more generic and fundamental.
Besides, SmartCheck~\cite{tikhomirov2018smartcheck}, Slither~\cite{feist2019slither}, eTainter~\cite{ghaleb2022etainter} and Madmax~\cite{grech2018madmax} tend to cover DoS vulnerabilities which are also caused by state-reverting. 
However, DoS is merely a sub-type of SRV, and the detection in eTainter~\cite{ghaleb2022etainter} relies on specific heuristics such as detecting gas insufficiency or exhausting. Such a design is not sufficiently effective and generic to all SRVs (i.e., unauthorized state changes).

\section{Conclusion}
\label{sec:conclusion}
This paper proposes \system{}, a static analysis framework for identifying state-reverting vulnerability in smart contracts. \system{} can be mainly divided into two procedures. Firstly, \system{} extracts the state dependency from the bytecode and historical transactions of smart contracts, and further merges the state dependency as a fine-grained state dependency graph. Secondly, \system{} utilizes a unique vulnerability detection method based on taint analysis guided by our proposed indicator analysis to find out the state-reverting vulnerability on the state dependency graph. We evaluate \system{} over a manually labeled dataset of 47 smart contracts and a large-scale dataset of 47,398 real-world smart contracts. The evaluation result shows that \system{} is effective in detecting state-reverting vulnerability with a high precision of 87.23\% and recall of 89.13\%. Further, SRVs are prevalent in real-world smart contracts. Particularly, we find that 11 SRVs exist in frequently-used smart contracts, affecting a total asset of 428,600 USD.

\section* {Acknowledgements} 
\label{sec:Acknowledgements}
The research was supported by the National Natural Science Foundation of China (62032025, 62202510), Special Projects in Key Fields of Universities in Guangdong Province (No. 2022ZDZX1001), the Fundamental Research Funds for the Central Universities, Sun Yat-sen University (22lgqb26), and the WeBankScholars Program.

\normalem


{
    \bibliographystyle{ACM-Reference-Format}
    \balance
    \bibliography{bib_bak}
}


\end{document}